\documentstyle[epsf]{mn}

\title[CDM models with a BSI steplike primordial spectrum and $\Lambda$]
{CDM models with a BSI steplike primordial spectrum and a cosmological
constant}
\author[J.~Lesgourgues, D.~Polarski and A.~A.~Starobinsky]
{Julien Lesgourgues$^{1}$, D. Polarski$^{1,2}$ 
and A. A. Starobinsky$^a$\\
$^1$~{\it Laboratoire de Math\'ematiques et de Physique Th\'eorique, 
UPRES-A 6083 CNRS}\\
{\it Universit\'e de Tours, Parc de Grandmont, F-37200 Tours (France)}\\
$^2$~{\it D\'epartement d'Astrophysique Relativiste et de Cosmologie},\\
{\it Observatoire de Paris-Meudon, 92195 Meudon cedex (France)}\\
$^a$~{\it Landau Institute for Theoretical Physics}, \\
{\it Kosygina St. 2, Moscow 117334 (Russia)}}

\date{\today}

\begin{document}

\maketitle

\begin{abstract}
A class of spatially-flat models with cold dark matter (CDM), 
a cosmological constant and a broken-scale-invariant (BSI) steplike
primordial (initial) spectrum of adiabatic perturbations, generated in an 
exactly solvable inflationary model where the inflaton potential has a rapid
change of its first derivative at some point, is confronted with existing 
observational data on angular fluctuations of the CMB temperature, galaxy
clustering and peculiar velocities of galaxies. If we locate the step
in the initial spectrum at $k\simeq 0.05
\,h\,\,{\mathrm Mpc}^{-1}$, where some feature in the spectrum of Abell
clusters of galaxies was found that could reflect a property of the initial 
spectrum, and if the large scales flat plateau of the spectrum is normalized
according to the COBE data, the only remaining parameter of the spectrum
is $p$\ - the ratio of amplitudes of the metric perturbations between the 
small scales and large scales flat plateaux.
Allowed regions in the plane of parameters ($\Omega=1-\Omega_\Lambda$, $H_0$)
satisfying all data have been found for $p$\ lying in the region (0.8-1.7). 
Especially good agreement of the form of the present power spectrum in this
model with the form of the cluster power spectrum is obtained for the 
{\em inverted} step ($p<1$, $p=0.7-0.8$), when the initial spectrum 
has slightly more power on small scales.
\end{abstract}


\section{INTRODUCTION}

The inflationary paradigm (see the review in Linde 1990; Kolb \& Turner 1990) 
offers an elegant solution to some 
of the outstanding problems of standard Big Bang cosmology. In these models, 
primordial quantum fluctuations 
(Hawking 1982, Starobinsky 1982, Guth \& Pi 1982) of some scalar field(s) 
(inflaton(s)) are produced, which eventually form galaxies, 
clusters of galaxies and the large-scale structure 
of the Universe through gravitational 
instability. Though there is a variety of possible models,
parametrized by a few number of constants, 
the increasing amount of data, for example from redshift surveys and
cosmic microwave background (CMB) 
anisotropies measurements, severely constrain the proposed models. Hence 
some of them can be definitely excluded at this stage while the remaining 
ones are found to be viable only in some well defined region of their free 
parameter(s) space. Additional sharp constraints are expected from the planned 
satellite missions MAP and PLANCK SURVEYOR for the measurement of the CMB 
anisotropies up to small angular scales.    

Since it is known that the simplest CDM model with a flat (n=1) initial
spectrum of adiabatic perturbations does not agree with observational data
(if normalized to the COBE data at large scales, it has too much power at 
small scales), a number of approaches to increase the ratio of large to
small scale power were proposed. One possibility is to change the initial
spectrum of perturbations. Since tilted scale-free spectra ($n<1$) did not
appear successful, the next step was to consider broken-scale-invariant (BSI)
spectra arising in inflationary models with two effective scalar fields
(Kofman, Linde \& Starobinsky 1985;
Kofman \& Linde 1987;
Silk \& Turner 1987;
Kofman \& Pogosyan 1988;
Gottl\"{o}ber, M\"{u}ller \& Starobinsky 1991;
Polarski \& Starobinsky 1992). 
Recently, the CMB anisotropies 
for a model of double inflation (Lesgourgues \& Polarski 1997) 
was investigated and it was found 
that for values of the parameters which yield a power spectrum $P(k)$ 
in fair agreement with observations, the Doppler peak turns out to be low. 
This is related to the effective tilt of the spectrum on very large scales.

Another possibility is to add a positive cosmological constant leaving the 
initial $n\simeq 1$\ spectrum unchanged. It was long known that the 
cosmological constant is viable only if it is accompanied by cold dark matter
and, vice versa, its inclusion improves the CDM model a great deal 
(as emphasized e.g. in Kofman \& Starobinsky (1985)).This model remains
viable after the detection of the CMB anisotropies on large 
angular scales by COBE (Kofman, Gnedin \& Bahcall 1993), 
and it is now perhaps the most 
promising CDM variant (Bagla, Padmanabhan \& Narlikar 1995;
Ostriker \& Steinhardt 1995). As well known, one important motivation 
for a positive cosmological constant $\Lambda$ is that it provides the 
possibility to accomodate both a high Hubble ``constant'', $h>0.6$, and a 
sufficiently old universe, $t_0>11$ Gyrs. Also, the baryon fraction in 
clusters seems to imply $\Omega\le 0.55$\ 
($\Omega=1-\Omega_\Lambda$\ stands for the total matter density including
CDM and baryons).
The most recent strong argument in favour of $\Omega< 1$\ 
(and $\Omega \simeq (0.2-0.4))$\ follows from the evolution
of rich galaxy clusters (Bahcall, Fan \& Cen 1997, 
see also Fan, Bahcall \& Cen 1997).

Up to now, these two possibilities were considered as mutually exclusive.
Now we want to unite them and compare the BSI CDM model including a 
cosmological constant with the observational data. The reasons for this are
the following. First,
it can enlarge the allowed cosmological parameters window. Second, 
the possibility exists that the initial power spectrum of scalar
(density) perturbations in the Universe is not scale free but has instead 
some non-trivial structure near $k=0.05 \,h\,\,{\mathrm Mpc^{-1}}$.
In fact observations may point to such a feature: the analysis of the 
three-dimensional distribution of 
rich Abell galaxy clusters located in superclusters, 
performed in Einasto {\it et al.} (1997a), 
pleads for an unexpected
spatial quasi-periodicity of the data
(see also Einasto {\it et al.} 1997b, 1997c).
Also, the spatial distribution of all Abell clusters of galaxies has
a well-marked
peak in the power spectrum at $k\simeq 0.05\,h\,\,{\mathrm Mpc^{-1}}$
(Einasto {\it et al.} 1997a). 
The Fourier power spectrum of the
spatial distribution of APM galaxies also has a feature on the same scale, 
though of a slightly different form (Caztanaga \& Baugh, 1997). Note that the
natural attempt to explain this feature by Sakharov oscillations
produced by the baryon admixture to CDM does not work 
(Atrio-Barandela {\it et al.}, 1997, Eisenstein {\it et al.}, 1997).
So this feature, if confirmed by future improved large scale structure
observations, should be ascribed to the initial perturbation spectrum itself.

Therefore, we need an initial spectrum which has a non-trivial structure
around some scale (preferably, with a bump) and has essentially no tilt
at larger and smaller scales. The latter condition is necessary in order to 
have sufficiently early galaxy and quasar formation. On the other hand, this
spectrum should be derivable from some first principles 
(e.g., it could be generated in a concrete inflationary model).
Such a spectrum naturally arises in a well-defined and rather generic 
(though idealized, of course) inflationary model where the inflaton potential
$V(\varphi)$\ has a 
local steplike feature in the first derivative. 
An exact analytical expression for the scalar (density) perturbations 
generated in this model was found in Starobinsky (1992).
It has a universal shape depending on only one parameter $p$.
Actually, it seems to be the only example of a perturbation spectrum with
the desired properties, for which a closed analytical form exists.

Thus, we suppose that the inflaton potential $V(\varphi)$\ has a 
rapid change of slope in a neighborhood $\Delta \varphi$\ of
$\varphi_0$:
\begin{eqnarray}
V(\varphi)
&=&V_0+v(\varphi), \\
v(\varphi)
&\simeq& A_+\varphi,\quad
\varphi>\varphi_0,\quad |\varphi-\varphi_0|\gg\Delta\varphi,
\nonumber\\
&\simeq& A_-\varphi,\quad 
\varphi<\varphi_0,\quad |\varphi-\varphi_0|\gg\Delta\varphi, \\
v(\varphi_0)&=&0, \quad A_+>0, \quad A_->0. \nonumber
\end{eqnarray}
The resulting adiabatic perturbation spectrum is non-flat around
the point $k_0=a(t_0) H(t_0)$, $t_0$\ being the time at
which $\varphi=\varphi_0$ while $H\equiv \dot{a}/a$ is the Hubble parameter.
One can show (Starobinsky 1992) 
that if the width $\Delta \varphi$\ of the singularity
is small enough, namely, 
$\Delta\varphi\,\,H(t_0)^2\ll\min(A_+,|A_+-A_-|)$, 
then the adiabatic perturbation spectrum has maximal deviation from
flatness, 
and acquires a universal form that can be derived analytically:
\begin{eqnarray} \label{Phi}
k^3 \Phi^2(k)\propto 1
-3(p-1)\frac{1}{y}\left( \left(1-\frac{1}{y^2} \right) \sin2y
+\frac{2}{y}\cos2y\right) \nonumber\\
+\frac{9}{2}(p-1)^2\frac{1}{y^2} \left(1+\frac{1}{y^2}\right)\times 
\nonumber\\
\left(1+\frac{1}{y^2}+\left(1-\frac{1}{y^2}\right) \cos2y
-\frac{2}{y}\sin 2y\right) \\
y=\frac{k}{k_0},  \qquad p=\frac{A_-}{A_+}, \nonumber
\end{eqnarray}
where $\Phi$\ is the (peculiar) gravitational potential.
This expression, plotted in Fig. \ref{figPHI}, 
depends (besides the overall normalization) 
on two parameters $p$\ and $k_0$.
The shape of the spectrum does not depend on $k_0$, $k_0$\ only determines
the location of the step. 
For $p>1$, the spectrum has a flat upper plateau on larger scales, 
even with a small bump, 
and a sharp decrease on smaller scales, with large oscillations though.
For $p<1$ this picture is inverted.  
The ratio of power between the plateaux equals $p^2$, and 
for $p=1$ we just recover the (flat) scale-invariant Harrison-Zel'dovich 
spectrum. 
Note that this spectrum cannot be obtained in the slow-roll approximation
(even with any finite number of adiabatic corrections to it).
In this model it is still possible to fix freely the amount of 
primordial gravitational waves (GW's) for given $p$ and normalization and we 
consider here the model with no GW's at all.
Without the inclusion of a cosmological constant, we would be forced to 
consider the case $p>1$\ only, in order to increase power on large scales. 
Since
$\Lambda>0$\ already produces a desired excess of large-scale power,
we are now free to consider both cases $p>1$\ and $p<1$.

We use for this study observational constraints both on the matter 
power spectrum $P(k)$ on one hand, and on the CMB anisoptropies on large, 
intermediate and small angular scales on the other hand, as done in 
Lesgourgues \& Polarski (1997), 
and we refer the interested reader to this article for more details.  

\section{CONFRONTATION WITH OBSERVATION}

In this work we restrict ourselves to 
the case of a spatially flat universe,
containing cold dark matter, baryonic matter with $\Omega_B h^2=0.015$, and 
a cosmological constant $\Lambda$. Hence, the
two cosmological parameters $h=H_0/100$\ and $\Omega_\Lambda$ are free. 
The present power spectrum $P(k)$\ reads:
\begin{equation}
P(k)=\frac{4}{9} \frac{k^4}{H_0^4} \Phi^2(k) T^2(k) \Omega^{-2}
(\frac{5}{3})^2 \left(1-\frac{H}{a}\int_0^t a\,dt \right)_{t=t_0}^2 ,
\end{equation}
where $\Phi(k)$, given by Eq. (3), is the gravitational potential
at the matter dominated stage for large redshifts $z\gg1$\ when 
$\Omega \simeq 1$, and we put the light velocity $c=1$.
The scale factor $a(t)$\ is given by the expression 
$a(t)=a_1 (\sinh (\frac{3}{2} H_{\infty} t))^{2/3}$\ where 
$H_{\infty}=\sqrt{\Lambda/3}=H_0 \sqrt{1-\Omega_{\Lambda}}$, $a_1=$const.
The transfer function $T(k)$\ is computed with the fast Boltzmann code  
{\sc cmbfast} by Seljak \& Zaldarriaga (1996), for each value of the
cosmological parameters. 

The power spectrum is normalized to the four years COBE DMR data 
(Bennett {\it et al.} 1996), using $Q_{{\mathrm rms-ps}|n=1}$\
(in the relevant cases, COBE scales will always correspond to the small-$k$ 
flat plateau of the initial spectrum). 
Afterwards, we use the following tests to discriminate between 
each set of values of the cosmological, resp. inflationary, parameters 
$h$, $\Omega_\Lambda$, resp. $p$, $k_0$):

\begin{enumerate}
\item
The ``optical'' $\sigma_8$. White, Efstathiou and Frenk (1993) give
$\sigma_8=(0.57\pm0.06)\,\, \Omega^{-0.56}$, with conservative errorbars.
This is a sharp constrain at wavenumbers 
$k\sim0.2\,h\,\,{\mathrm Mpc^{-1}}$. 
More recent determinations of this quantity have a tendency to decrease
it to $\sigma_8\sim 0.5$, and even a bit lower (Ebe, Cole \& Frenk 1996;
Viana \& Liddle 1996; Ying, Mo \& B\"{o}rner 1997).
Still, we shall use the former value (the exponent of $\Omega$
corresponds to the case of a flat Friedmann-Robertson-Walker model
with a $\Lambda-$term, see below).

\item
Peculiar velocities, deduced from the Mark III catalog, 
and {\sc POTENT} reconstruction of the density field.
In our case, the power spectrum has got strong oscillations, so
we cannot simply use direct estimates of $P(k)$\ at given 
wavenumbers (Kolatt \& Dekel 1997),
which would give precise constraints in the case of a smooth spectrum.
We will rather use the rms bulk velocity in a sphere
of radius $R$:
\begin{equation}
\langle V_R^2 \rangle =
\frac{f^2(\Omega) H_0^2}{2\pi^2}\int_0^\infty dk P(k) \tilde{W}^2_R(k),
\end{equation}
where $\tilde{W}_R(k)$\ is the Fourier transform of the top-hat window 
function of radius $R$, and $f(\Omega) \equiv H^{-1} \dot{D}/D$\
($D(t)$\ is the linear growth factor for inhomogeneities).
Expressing $f$\ as a power law, $f(\Omega)=\Omega^{r}$, one can easily
compute the index for a given $\Omega$\ and $\Omega_\Lambda$. 
In the interesting range 
$0.2\leq \Omega \leq 1$, $r=0.57-0.60$\ for an open universe with 
$\Omega_\Lambda=0$, but $r=0.55-0.56$\ for a flat 
universe with $\Omega+\Omega_\Lambda=1$. In the following we will take
$f(\Omega)=\Omega^{0.56}$. 
The Mark III POTENT result at $R=50h^{-1} {\mathrm Mpc}$\ (with a gaussian
smoothing at $R_s=12 h^{-1} {\mathrm Mpc}$) is 
$V_R=375\pm85 \ {\mathrm km\ s}^{-1}$ (Kolatt \& Dekel 1997). 
The cosmic variance 
(the possible dissimilarity between the rms value of $\langle V_R^2 \rangle$ 
and the particular realization in our local neighborhood) is quite large for 
this quantity ($\sim100\ {\mathrm km\ s}^{-1}$), 
and can be added in quadrature with the previous
errorbar, leading to a global uncertainty 
$\sigma \simeq 130\ {\mathrm km\ s}^{-1}$. This test is mainly sensitive to 
wavenumbers 
$0.01 \leq k\leq 0.06\,h\,\,{\mathrm Mpc^{-1}}$.

\item
Redshift surveys. Since they are strongly bias-dependent, redshift
surveys give indications about the shape of $P(k)$. 
Here again, due to the oscillations, 
instead of using some sets of estimates at given wavenumbers, 
one has to convolve the spectrum with the 
window functions of a given experiment and compare with the raw data.
We use the count-in-cells analysis of large-scale
clustering of the Stromlo-APM redshift survey. Taking other experiments
into account would slightly improve the precision, but not change the
results, since Stromlo-APM is in very good agreement with other redshift 
surveys, as can be seen in Peacock \& Dodds (1994). 
After normalizing the spectrum to $\sigma_8=1$, we compute the variance
$\sigma_l^2$\ in cells of size $l\,h^{-1}{\mathrm Mpc}$, and
compare it with the data (Loveday {\it et al.}, 1992), 
consisting
of nine points (assumed to be independent, with error bars treated 
as $2\sigma$ ones), through a $\chi^2$\ analysis. Since we can vary
four parameters (plus the overall normalisation,
which is irrelevant for this test), 
$\chi^2 \leq 5$\ is excellent,
whereas $\chi^2 \geq 15$ is bad.

\item
CMB anisotropies.  
We compute the curves $l(l+1)C_l$ using {\sc cmbfast}, and compare it with
some preliminary measurements. At the moment, there are still many 
uncertainties, and we only have global indications on the $C_l$'s curve.
As far as the first peak is concerned, 
a sixth order polynomial fit to the full available data set gives 
$A_{peak}\equiv [l(l+1)C_l/2\pi]^{1/2}=28\times 10^{-6}~{\rm with}~l=260$
(Lineweaver \& Barbosa, 1997),
but it is very difficult to calculate 
an errorbar for this quantity in the general case. 
CAT and OVRO give an indication on the amount of power on small scales,
but do not constraint the position and height of the secondary peaks.

Since the precision of these measurements is increasing very quickly, 
we do not intend in this work to perform a full $\chi^2$\ analysis, 
using each result and the corresponding window function 
(to find which parameters yield the best agreement).
We prefer to calculate the $C_l$'s and comment
our results in such way that in a few years one could 
easily update the analysis,
restrict the allowed parameters window and eventually rule out the model. 
This is why we concentrate
on the position and height of the peaks. 
We will use CMB data to eliminate
parameters only if there is an obvious discrepancy between 
the predicted curve and the observations.
 
\end{enumerate}

\section{RESULTS}

Starting with a flat spectrum, we explore the range
$h=0.5,0.6,0.7$\ and $0\leq\Omega_\Lambda\leq 1$. 
For each value of $h$, there are some $\Omega_\Lambda$'s in agreement with 
the $\sigma_8$\ constraint:
($h=0.5$, $0.45\leq\Omega_\Lambda\leq0.50$), 
($h=0.6$, $0.55\leq\Omega_\Lambda\leq0.60$) and
($h=0.7$, $0.65\leq\Omega_\Lambda\leq0.70$).
These windows are inside the limits
$\Omega h \simeq 0.25-0.30$\ found in Kofman, Gnedin \& Bahcall (1993),
and are in good agreement with other tests: bulk velocity, with
$V_{50} \simeq 300\ {\mathrm km\ s}^{-1}$, 
and count-in-cells, with $3<\chi^2<6$. There is
no obvious contradiction with CMB measurements, and the first Doppler peak
is fairly high: $A_{peak}\simeq (26-29) \times 10^{-6}$.
Therefore, it is not necessary to depart from a flat spectrum in order to
explain all observations (apart, of course, from a possible spike 
in the spectrum at $k\approx0.05 \,h\,\,{\mathrm Mpc}^{-1}$) 
if $h$\ and $\Omega_\Lambda$\ turn out to be
close to these values.

However, as we shall see, the steplike spectrum is 
compatible with a larger subset ($h$, $\Omega_\Lambda$).
It also predicts some specific features in $P(k)$\ and $C_l$ curves 
that could easily be observed or ruled out by future experiments, 
so we are {\it not} just adding some extra degeneracy.
More precisely, one can think of a spectrum with:
\begin{description}
\item[\rm A.]
$p>1$, in order to get less power on small scales in the primordial spectrum.
To compensate the loss of power in $P(k)$, we will allow smaller values 
of the cosmological constant. This will lower the CMB 
anisotropies. Then, to avoid a problematic collapse of the first acoustic 
peak, like in double inflation (Lesgourgues \& Polarski, 1997), 
$k_0$\ must be chosen so that multipoles up to
$l\sim200$\ (at least) are given by the upper plateau of the primordial 
spectrum. This means that we can forget any
$k_0 < 0.03 \,h\,\,{\mathrm Mpc^{-1}}$. 
For $k_0 \simeq 0.03 \,h\,\,{\mathrm Mpc^{-1}}$,  
the first peak is even enhanced by a few percents by the global maximum 
of the primordial spectrum 
(the ``bump'' at the extremity of the upper plateau). 
\item[\rm B.]
$p<1$, in order to get more power on small scales in the primordial spectrum,
and therefore allow some higher values of the cosmological constant 
which would be excluded by small scale constraints (for instance, $\sigma_8$) 
in case of a flat spectrum ($p=1$).
Since CMB peaks grow with $\Omega_{\Lambda}$,
multipoles $l\gg 2 k_0/a_0 H_0$\ will be unusually large.
A priori, in this case, the step is anywhere between the COBE and 
$\sigma_8$\ scales: 
$0.003 \leq k_0 \leq 0.1\,h\,\,{\mathrm Mpc^{-1}}$.
However, in this case the spectrum (3) has a well-pronounced
sharp maximum at $y\simeq 3.5$\ for the values $p\simeq 0.8$\
which are the most interesting ones as will be seen below (for $p\ll 1$
the maximum is located at $y\simeq 3.14$).
So, if we want to use this bump to explain the feature in the cluster spectrum
at $k \simeq 0.05 \,h\,\,{\mathrm Mpc^{-1}}$, $k_0$\ should be taken
$\approx 0.015 \,h\,\,{\mathrm Mpc^{-1}}$.
Note that this possibility was not expected and discussed before.
\end{description}

\subsection{The case $p>1$}

We consider first the case $h=0.5$. 
Assumption A turns out to be successfull 
with respect to the first three tests in many cases: 
for any $0\leq \Omega_{\Lambda} \leq 0.5$, one can find a large allowed 
window in the $(p,k_0)$\ plane. Of course, a smaller $\Omega_{\Lambda}$\
will lead to a higher range for $p$. For instance, when 
$\Omega_{\Lambda}=0.3$, we find $1.3\leq p\leq 1.7$\ and 
$0.03 \leq k_0 \leq 0.06\,h\,\,{\mathrm Mpc^{-1}}$. 
We must stress that satisfying the three tests
is a success, since most other models satisfy two of them at most.
For instance, if a tilted spectrum leads to a correct $\sigma_8$\
and $\chi^2$, the bulk velocity will be generally too low, unless a very large
cosmic variance is invoked. Double inflation will
predict a higher $V_{50}$, but still under the $-1\sigma$\ errorbar. 
In the present model, $V_{50}$\ is much higher, in very good agreement with 
observations, as can be seen on table I (second line) for one example,
because there is at least as much power on scales
$0.01 \leq k \leq 0.06\,h\,\,{\mathrm Mpc^{-1}}$\ 
as for a scale invariant spectrum.
    
Including  CMB anisotropies in the tests
provides two independent constraints: 
\begin{itemize}
\item
on one hand, the position and height of the first peak are essentially
related to cosmological parameters, not inflationary parameters. 
Indeed, as we said previoulsly, 
in the relevant cases ($k_0\geq0.03 \,h\,\,{\mathrm Mpc^{-1}}$), 
the first peak is deduced from an essentially scale invariant spectrum 
(with only 
a little enhancement proportional to $p$, but $\leq 10 \%$, on
$A_{peak}$\ in viable cases), and depends only on $h$\ and 
$\Omega_{\Lambda}$. 

The position and height of the first peak are not precisely constrained 
by observations at the moment, neither by Saskatoon 
(whose calibration is under progress: Leicht, in preparation), 
nor by MSAM. 
Inside the allowed $(p,k_0)$\ window found previously, we find 
$24\times 10^{-6}< A_{peak}< 30\times 10^{-6}$, in good agreement with
current limits, so we cannot exclude any set of parameters.
\item  
on the other hand, experiments on the secondary peaks scales 
constrain $p$\ and $k_0$. The global height of the multipoles at
$400<l<1500$ depends on $p$, whereas $k_0$\ gives the detailed shape 
at these scales (for instance,
the ratio between the peaks), by shifting the maxima and minima of the 
primordial spectrum in $k-$space. 

At the moment, observations do not indicate a detailed
shape, but from CAT and OVRO we know that multipoles
on such scales should range basically between 
$A_l=[l(l+1)C_l/2\pi]^{1/2}=10\times 10^{-6}$\ and
$A_l=25\times 10^{-6}$. 
Since we are not dealing with the detailed 
window function of each measurement, we must be extremely conservative.
Using the second point of CAT, one can state that a $C_l$
curve that would not reach 
$A_l=12\times 10^{-6}$ 
(the $-1\sigma$\ value) in the range $550<l<720$\
(for which the window function is above half of its peak value) 
can be confidently excluded. The reason for which we use this particular point
is that the associated window function does not interfear with the first 
acoustic peak: it is probing power only on the scales of the secondary ones. 
This restriction provides, for each value of the cosmological constant,
an upper limit on $p$, and we find that for $0\leq \Omega_{\Lambda} < 0.2$,
all sets of parameters are ruled out. 
This is an indirect constraint on $\Omega_{\Lambda}$.
For $0.2 \leq \Omega_{\Lambda} \leq 0.5$, the previously found windows 
still hold. 
\end{itemize}
Finally, for $0.2 \leq \Omega_{\Lambda} \leq 0.5$, 
all the constraints can be satisfied by some values of $p$\ and $k_0$\ in
the ranges 
$1 \leq p \leq 1.7$\ 
and $0.03 \leq k_0 \leq 0.07\,h\,\,{\mathrm Mpc^{-1}}$.
To illustrate this case, we show in table I (second line) a particular
example: $h=0.5,\ \Omega_{\Lambda}=0.3,\ p=1.3,\ 
k_0=0.03 \,h\,\,{\mathrm Mpc^{-1}}$\ ($\Omega_{\Lambda}$\ is chosen to 
obtain the preferred value $A_{peak}=28\times 10^{-6}$, 
and $p$\ is as low as possible, 
in order to maximize small scales anisotropies).
We also give an example of the
case $\Lambda=0$, $p=2.1$, though it is excluded by CAT.
The corresponding power spectra are plotted in Fig. \ref{figPK},
the CMB anisotropies in Fig. \ref{figCMB}. 

This type of model could be easily discriminated by the 
forthcoming improvements
of redshift surveys and CMB observations. The former  
might state about the little well predicted in the $P(k)$\
around $k\simeq (0.1-0.2) \,h\,\,{\mathrm Mpc^{-1}}$. 
The latter will soon indicate:
\begin{itemize}
\item first, the position and amplitude of the first peak, i.e.,
$h$\ and $\Omega_\Lambda$\ (in the framework of this model).
\item second, power on small scales, i.e. $p$.
\item finally, the shape of secondary peaks, i.e. $k_0$.
\end{itemize}
This model is very unlikely to be degenerate with some other one
(for instance, other cosmological parameters plus tilted spectrum)
from the point of view of CMB anisotropies, because
it predicts a tremendously high ratio between multipoles at 
scales $l\sim200$\ and $l\sim600$
(recall that, in contrast with tilted, $n<1$, or with double inflationary 
models, small scales are lowered however
intermediate scales are preserved). 

A similar analysis can be performed for higher $h$\ values.
Since increasing $h$\ lowers the CMB multipoles, 
$p$\ is more restricted now by the constraints
on small scales anisotropies. 
At $h$=0.6, possible models have $0.4 \leq \Omega_{\Lambda} \leq 0.6$,
$1 \leq p \leq 1.5$\ and $0.03 \leq k_0 \leq 0.07\,h\,\,{\mathrm Mpc^{-1}}$. 
The first peak reaches lower values as well: $26 \times 10^{-6} \leq A_{peak} \leq 27\times 10^{-6}$. 
When $h=0.7$, we find $0.5 \leq \Omega_{\Lambda} \leq 0.7,~1\leq p \leq 1.4,~
0.03 \leq k_0 \leq 0.07~h~{\rm Mpc}^{-1}$ and furthermore 
$24.5\times 10^{-6} \leq A_{peak} \leq 26\times 10^{-6}$.
The resulting allowed region in the ($h$, $\Omega_\Lambda$) plane is plotted 
in Fig. \ref{figWINDOW}. For a few successfull examples, we give the results
of the tests in Table I.

\subsection{The case $p<1$}

Again, we first consider the case $h=0.5$. 
When $0.45\leq\Omega_{\Lambda}\leq0.7$, one can find some
$(p,k_0)$\ in good agreement with $\sigma_8$, $V_{50}$\ and $\chi^2$.
For instance, when 
$\Omega_{\Lambda}=0.6$, the allowed region is $0.75 \leq p \leq 1$\ and 
$0.003 \leq k_0 \leq 0.04\,h\,\,{\mathrm Mpc^{-1}}$. 
In the last subsection, it was found that for $p>1$, 
the most compelling constraint was on $\sigma_8$.
Now the three tests play an important 
part in the definition of the allowed region.
Indeed, the above mentioned sharp maximum appears in $P(k)$, 
preceeded at larger scales by a depression at $y\approx 1.2$\
(the inverted bump of the case $p>1$). As a result, the power
spectrum $P(k)$\ has no pronounced second maximum at the place where it 
exists for $p=1$, namely $k\simeq 0.05~\Omega^{-1}h~{\rm Mpc}^{-1}$ .
Note that this depression would become very pronounced in the case $p\ll1$,
its position in this limit being given by $y=\sqrt{2.5 p}$\ 
(Starobinsky, 1992).
When $k_0>0.015\,h\,\,{\mathrm Mpc^{-1}}$, 
the bulk velocity is sometimes too small due to
this little depression. 
On the contrary, when $k_0<0.015\,h\,\,{\mathrm Mpc^{-1}}$,
the maximum often generates excessive bulk velocities.

As expected, the CMB anisotropies are amplified by both the comological 
constant and the primordial spectrum step. The basic picture is that the
$C_l$'s are enhanced by a factor $p^2$\ for $l>2 k_0/a_0 H_0\simeq12\,000k_0$.
When $k_0=0.01-0.02\,h\,\,{\mathrm Mpc^{-1}}$, the
first peak is enhanced by the maximum of the primordial spectrum,
so its location and maximum value are highly dependent
on all parameters, including $k_0$\ and $p$\ (in contrast with the case 
$p<1$). The secondary peaks are given by an almost flat region of the 
primordial spectrum, so they depend on all parameters, $k_0$ excepted.

As in the previous subsection, 
we can use the last CAT point to reduce the allowed window, confidently 
excluding any $C_l$ curve that would not pass 
$A_l=21\times 10^{-6}$ (the $-1\sigma$\ value) in the range $550<l<720$.
This rules out many low $p$\ values for a given
$\Omega_{\Lambda}$. In fact models with $\Omega_{\Lambda}>0.5$\ do not survive.
At $\Omega_{\Lambda}=0.5$\ we find the allowed window: $0.85 \leq p \leq 1$,
$0.003<k_0<0.04\,h\,\,{\mathrm Mpc^{-1}}$.
Similarily, when $h=0.6$, successfull models can be found for 
$0.55\leq\Omega_{\Lambda}\leq0.65$, extending the validity range of 
the scale invariant model.
At $\Omega_{\Lambda}=0.65$\ the
allowed window is $0.80<p<0.85$, 
$0.003 \leq k_0 \leq 0.04\,h\,\,{\mathrm Mpc^{-1}}$. 
Finally, when $h=0.7$, $0.65\leq\Omega_{\Lambda}\leq0.75$\ is allowed.
At $\Omega_{\Lambda}=0.72$\ we find
$0.80<p<0.85$, $0.003 \leq k_0 \leq 0.04\,h\,\,{\mathrm Mpc^{-1}}$.
These results are also summarized in Fig. \ref{figWINDOW}. Table I
contains a few examples, and for one of them the power spectrum and
CMB anisotropies are illustrated in Fig. \ref{figPK} and Fig. \ref{figCMB}.  

At first sight, the case $p<1$\ is not interesting since
it does not extend very much the allowed region for
($h$, $\Omega_{\Lambda}$):
good results are obtained for cosmological parameters that are not in conflict
with the scale invariant model.
The interest of the $p<1$\ steplike spectrum lies in the prediction
of specific features, namely:
\begin{itemize}
\item a sharp maximum in $P(k)$. 
The steplike model with $k_0\simeq0.015\,h\,\,{\mathrm Mpc^{-1}}$\ 
and $p\simeq0.8$\ could perfectly explain the form of the cluster
spectrum with a peak
at $k_0\simeq0.05\,h\,\,{\mathrm Mpc^{-1}}$\ (see Fig. \ref{figPEAK}). 
\item large CMB anisotropies. The Saskatoon experiment (Netterfield 
{\it et al.} 1997) 
indicates a too high first peak that cannot be explained by current 
flat CDM models
(unless $h=0.2-0.3$\ is allowed). 
These measurements might be contaminated
by some systematic effects, as indicated by MSAM third
flight result (Cheng 1997). However, if the Saskatoon points are confirmed,
the flat $\Lambda+$CDM steplike model 
with $p<1$\ would be a good condidate, since it
predicts high anisotropies, without getting in conflict with
constraints on $P(k)$, and without requiring $h<0.5$. 
For instance, when $p=0.85$, the $C_l$ values increase by 
$40\%$ at the scales of secondary peaks, and even of the first peak if 
$k_0\leq0.01\,h\,{\mathrm Mpc^{-1}}$. When $h=0.6$, 
$\Omega_\Lambda=0.65$, $p=0.85$, and $k_0=0.01h\,\,{\mathrm Mpc^{-1}}$, 
we find $A_{peak}=35\times 10^{-6}$.
\end{itemize}

\begin{table*}
\begin{minipage}{120mm}
\caption{Results of the tests for the different models. For each value of 
$h$, we show the best model with a flat spectrum ($p=1$), a step towards
large scales ($p>1$), and a step towards small scales ($p<1$). 
For $h=0.5$\ we also give the best model with $\Omega_\Lambda=0$.}
\begin{tabular}{|c|c|c|c|c|c|c|c|c|c|}
$h$&$\Omega_\Lambda$ 
&$p$ &$k_0$ & $\Omega^{0.56}\sigma_8$ & $V_{50}$ & $\chi^2$  
& first peak & second peak \\
&&&$(h\,\,{\mathrm Mpc^{-1}})$ && ($\mathrm{km\ s}^{-1}$) & 
&$l_{peak}$, $A_{peak}$&$l_{peak}$, $A_{peak}$\\
\hline
0.5&0&2.1&0.030&0.63&390&7.6
&$215, \ 26\times 10^{-6}$&$475, \ 10\times 10^{-6}$\\
&0.3&1.3&0.030&0.63&345&6.2
&$225, \ 28\times 10^{-6}$&$515, \ 16\times 10^{-6}$\\
&0.5&1&(flat)&0.54&300&3.7
&$235, \ 29\times 10^{-6}$&$555, \ 22\times 10^{-6}$\\
&0.5&0.85&0.015&0.63&310&3.1
&$260, \ 32\times 10^{-6}$&$555, \ 25\times 10^{-6}$\\
\hline
0.6
&0.55&1.2&0.030&0.53&330&4.6
&$220, \ 27\times 10^{-6}$&$510, \ 18\times 10^{-6}$\\
&0.60&1&(flat)&0.54&305&3.5
&$225, \ 27\times 10^{-6}$&$530, \ 21\times 10^{-6}$\\
&0.65&0.8&0.015&0.57&300&2.6
&$255, \ 31\times 10^{-6}$&$530, \ 27\times 10^{-6}$\\
\hline
0.7
&0.60&1.2&0.030&0.57&340&4.7
&$210, \ 25\times 10^{-6}$&$485, \ 18\times 10^{-6}$\\
&0.65&1&(flat)&0.57&315&4.1
&$215, \ 26\times 10^{-6}$&$500, \ 20\times 10^{-6}$\\
&0.70&0.8&0.015&0.59&310&2.7
&$240, \ 28\times 10^{-6}$&$505, \ 26\times 10^{-6}$\\
\end{tabular}
\label{table}
\end{minipage}
\end{table*}

\section{CONCLUSIONS AND DISCUSSION}

We have compared the CDM$+\Lambda$\ cosmological model with a BSI
initial spectrum of adiabatic perturbations given by Eq. (3)
with recent observational data. The model is determined by four fundamental
parameters $\Omega_\Lambda$, $A_+$, $A_-\equiv p A_+$\ and $k_0$, (in 
addition to the Hubble constant $H_0$) out of which one ($A_-$) is
fixed by the normalization to the COBE data.
The number of observational tests 
we use is sufficient to rule out many primordial 
spectra well-motivated by inflationary theories. For instance, to enlarge
the allowed ($h, \Omega_\Lambda$) window, one could
think of introducing a tilted or double inflationary spectrum
to reconcile observations on large scales (COBE) and small scales
($\sigma_8$). However, there will be a generic lack of power on intermediate
scales (bulk velocity, first CMB peak). 
Moreover, exactly at these scales, there may be 
an unexpected excess of power.
The initial spectrum that we study here 
allows a significant enlargement of the allowed
($h, \Omega_\Lambda$) region, especially smaller
$\Omega_\Lambda$'s, without supressing power at intermediate scales.

We have found allowed regions in the ($h$, $\Omega_\Lambda$) parameter plane 
for $p$\ lying in the region $(0.8-1.7)$.
These allowed regions are larger than in the case of a flat initial 
spectrum ($p=1$).
The most interesting, and alltogether unexpected, successfull model 
appears to be that with an inverted step $p<1$, where
the power at intermediate scales is even more enhanced.
It appears that this latter case is suitable for the description
of the feature in the cluster spectrum found in 
Einasto {\it et al.} (1997a, 1997b, 1997c).
The most distinctive feature of the class of models in question
is the suppression of the second  and higher acoustic (Doppler) peaks 
in the case $p>1$,
and their enhancement in the opposite case. 
That is why the CAT CMB experiment appears the most restrictive for the model.
So, the exact measurement
of $C_l$\ for $l\sim500$,
i.e. around the second acoustic (Doppler) peak, 
will be the crucial test for this model.
The forthcoming improvements of CMB anisotropies
measurements, especially baloon and satellite experiments, should be able
either to rule out this model or to detect its signature in the 
next ten years.

On the other hand, the increase of the allowed region in the 
($h$, $\Omega_\Lambda$)
plane and the allowed range for $p$\ itself are not large. This shows
the remarkable robustness of the CDM$+\Lambda$\ cosmological model
with the simplest inflationary initial conditions 
($\Omega_{\mathrm tot}=\Omega+\Omega_\Lambda=1$; $n\simeq 1$).
Also, the fact that the allowed values of $p$\ are close to unity indicates
that the form of the inflaton potential $V(\varphi)$\ is close to the
case of a discontinuity in its second, not first derivative, which is more
natural since, e.g., it can occur as a result of an equilibrium
second-order phase transition (some kind of non-analytic behaviour of
$V(\varphi)$\ is required in any case to obtain significant deviations
from the flat perturbation spectrum). Consideration of the latter case is 
under progress.  

\section*{Acknowledgements}
A.S. thanks the Ecole Normale Sup\'{e}rieure, Paris, 
for financial support, under the agreement between the Landau Institute for
Theoretical Physics and ENS, during his visit to France when this
paper was completed.
A.S. also acknowledges financial support 
by the Russian Foundation for Basic Research, 
grant 96-02-17591, and by the Russian research project
``Cosmomicrophysics''.

\begin{figure*}

\epsfysize=6.5cm

\caption{\small Primordial power spectrum for 
($p=0.8$, $k_0=1\,h\,\,{\mathrm Mpc}^{-1}$) and 
($p=1.7$, $k_0=1\,h\,\,{\mathrm Mpc}^{-1}$).
At this stage, the normalisation of each spectrum is arbitrary.
It is important to note that in both cases, the maximum 
is located at the extremity of the upper plateau.} 
\label{figPHI}
\end{figure*}

\begin{figure*}

\epsfsize=6.5cm

\caption[]{\small Allowed region in the cosmological parameters plane
($h$, $\Omega_{\Lambda}$). The lower hatched region corresponds to models
with $p\geq1$, the upper one to models with $p\leq1$. 
Inside the intersection, a scale-invariant spectrum ($p=1$) is allowed.
The steplike model is seen to enlarge significantly the allowed region.} 
\label{figWINDOW}
\end{figure*}

\begin{figure*}

\epsfsize=10.5cm

\caption[]{\small Power spectrum for a few models from Table \ref{table} :
a model with $\Omega_\Lambda=0$\ (in conflict with CAT), 
and three viable models with
$p>1$, $p=1$\ and $p<1$.} 
\label{figPK}
\end{figure*}

\begin{figure*}

\epsfsize=10.5cm

\caption[]{\small CMB anisotropies for the same models as in the previous 
figure.
We also plot a few measurements, including
Saskatoon (Netterfield 1997) recalibration (Leicht, in preparation) and
new preliminary CAT (Baker, in preparation) 
and OVRO (Leicht, in preparation) results. 
We have in order of appearance for growing $l$: COBE (3 points), Tenerife, 
South Pole, Saskatoon (5 points), MAX (2 points), MSAM, CAT (2 points) and 
OVRO.} 

\label{figCMB}
\end{figure*}

\begin{figure*}


\caption[]{\small Theoretical power spectrum for $h=0.7$, 
$\Omega_\Lambda=0.72$, $p=0.75$\ ans $k_0=0.016\,h\,{\mathrm Mpc}^{-1}$\
compared with the power spectrum of rich Abell galaxy clusters, taken from
Einasto {\it et al.} (1997a) and divided by $b^2=5$.} 

\label{figPEAK}
\end{figure*}

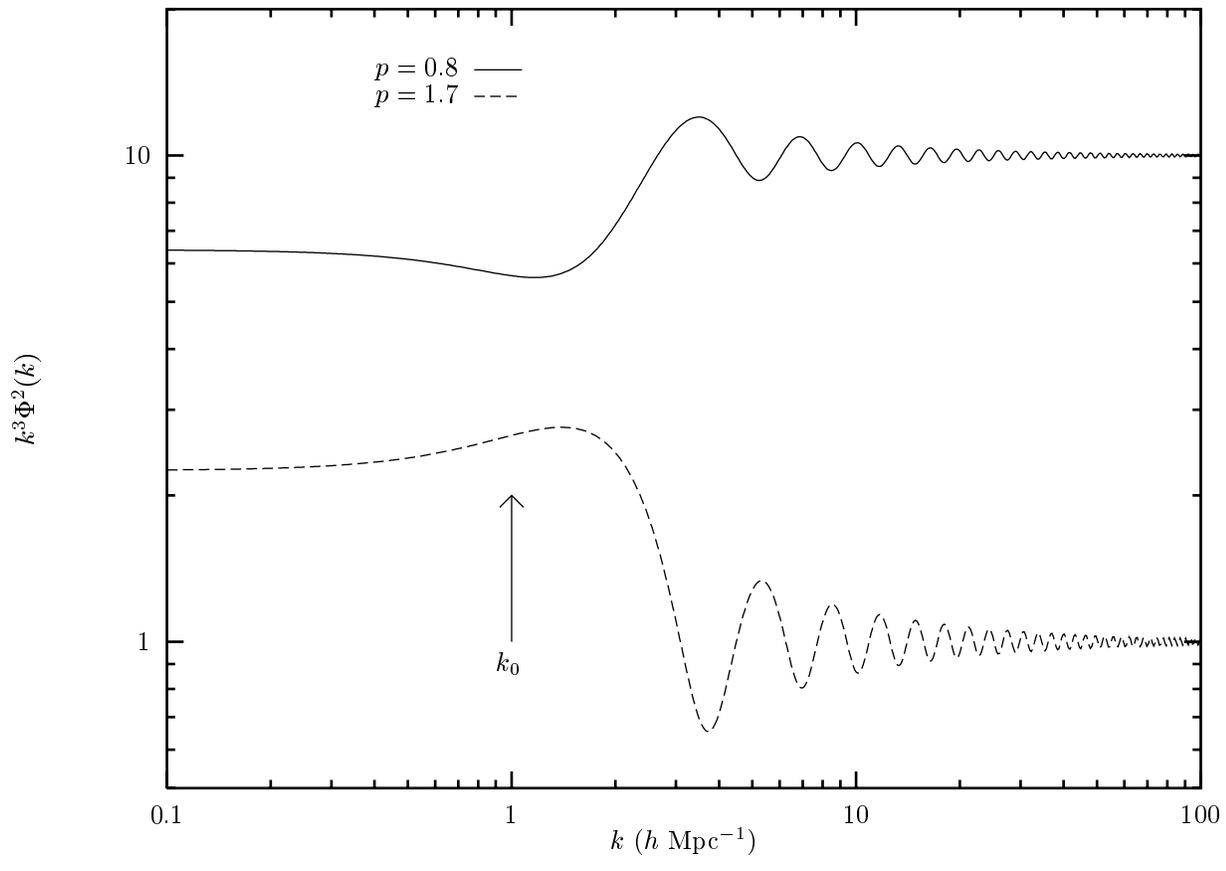
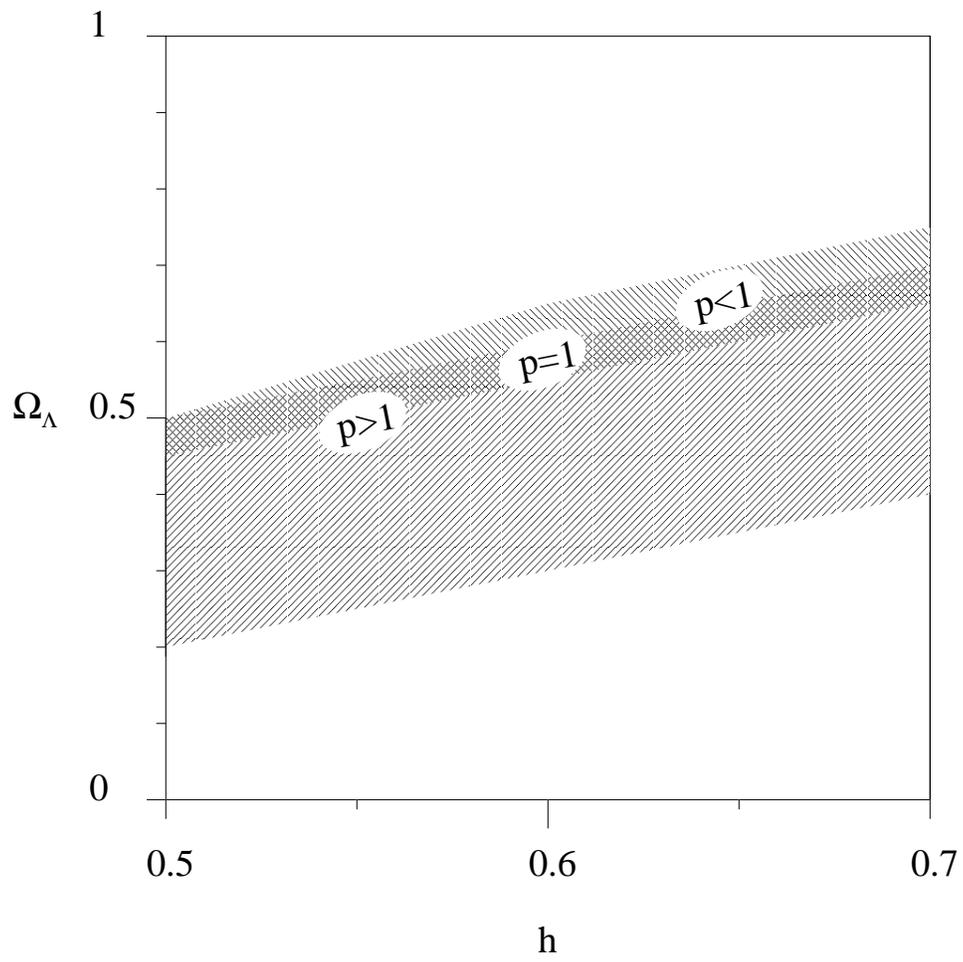
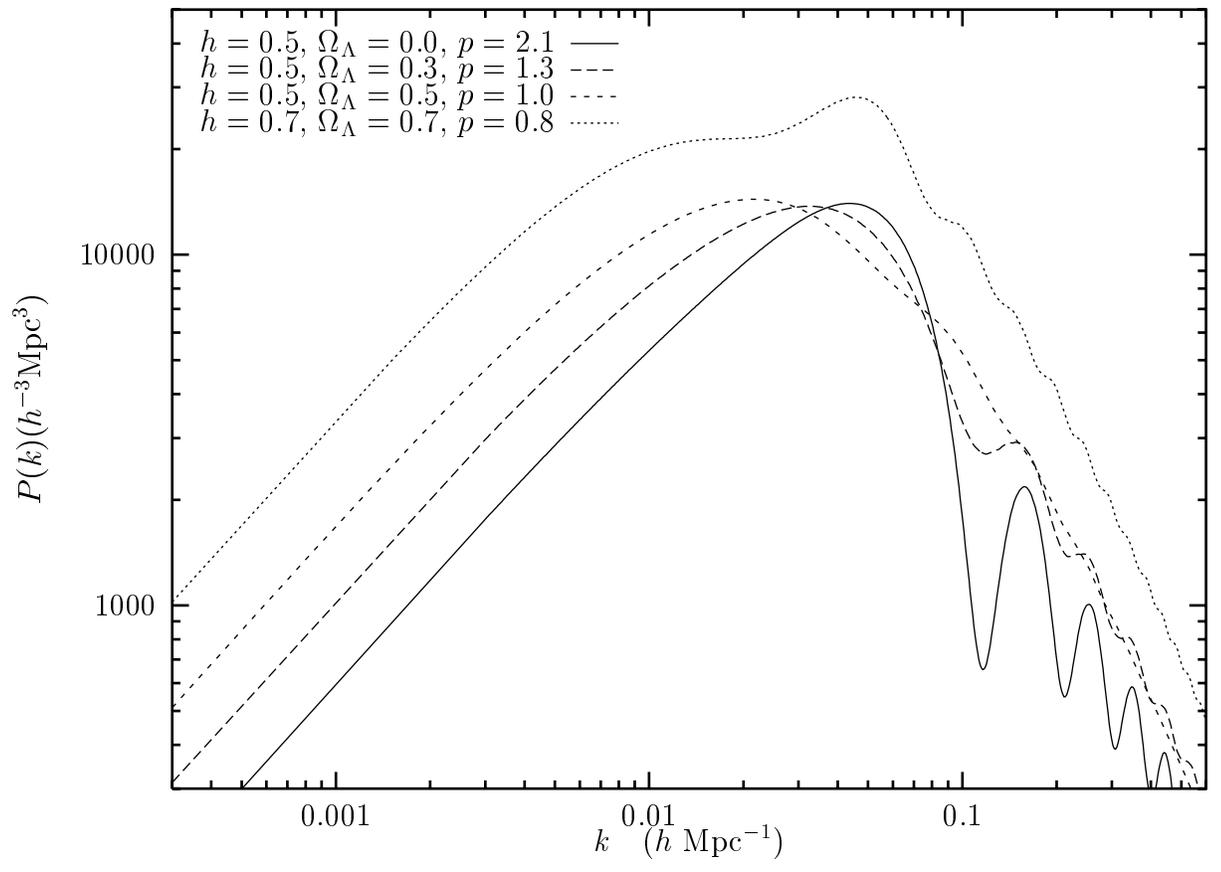
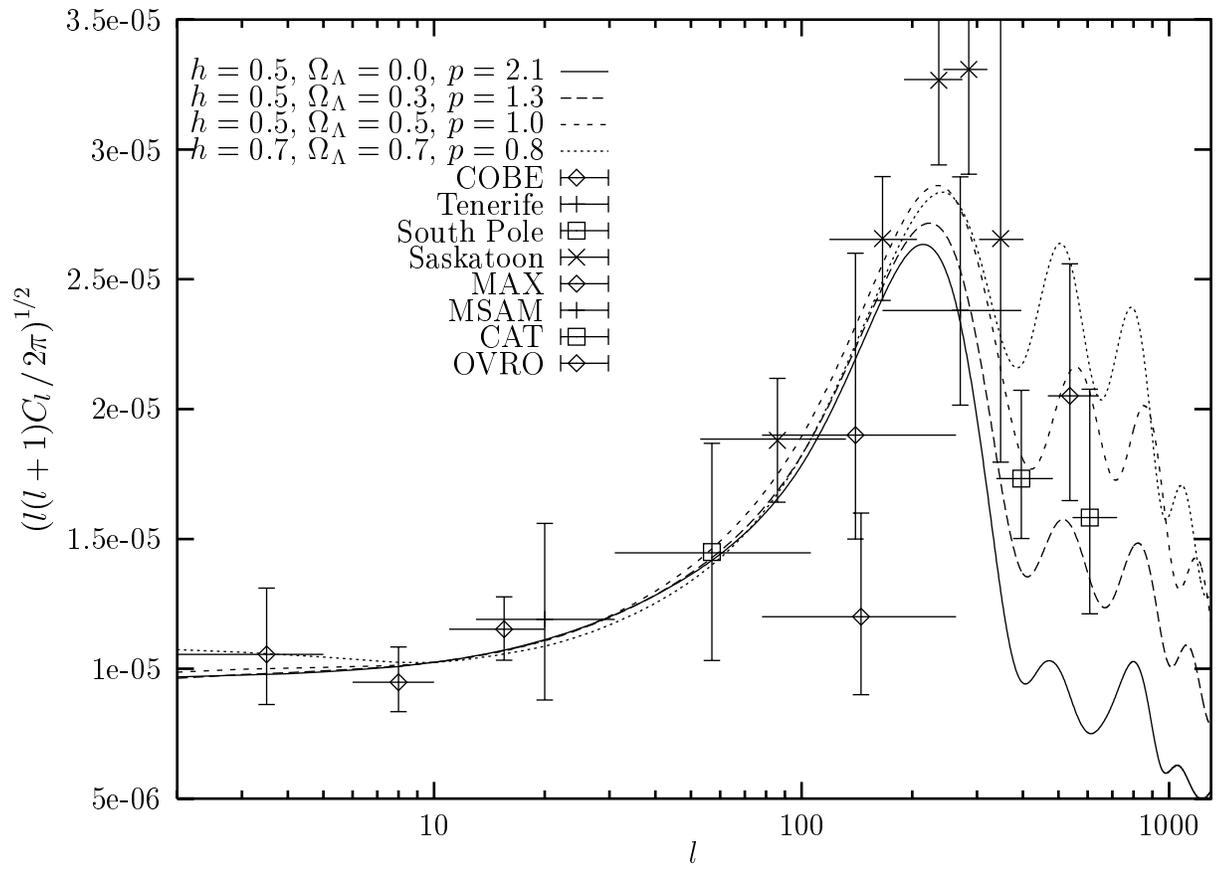
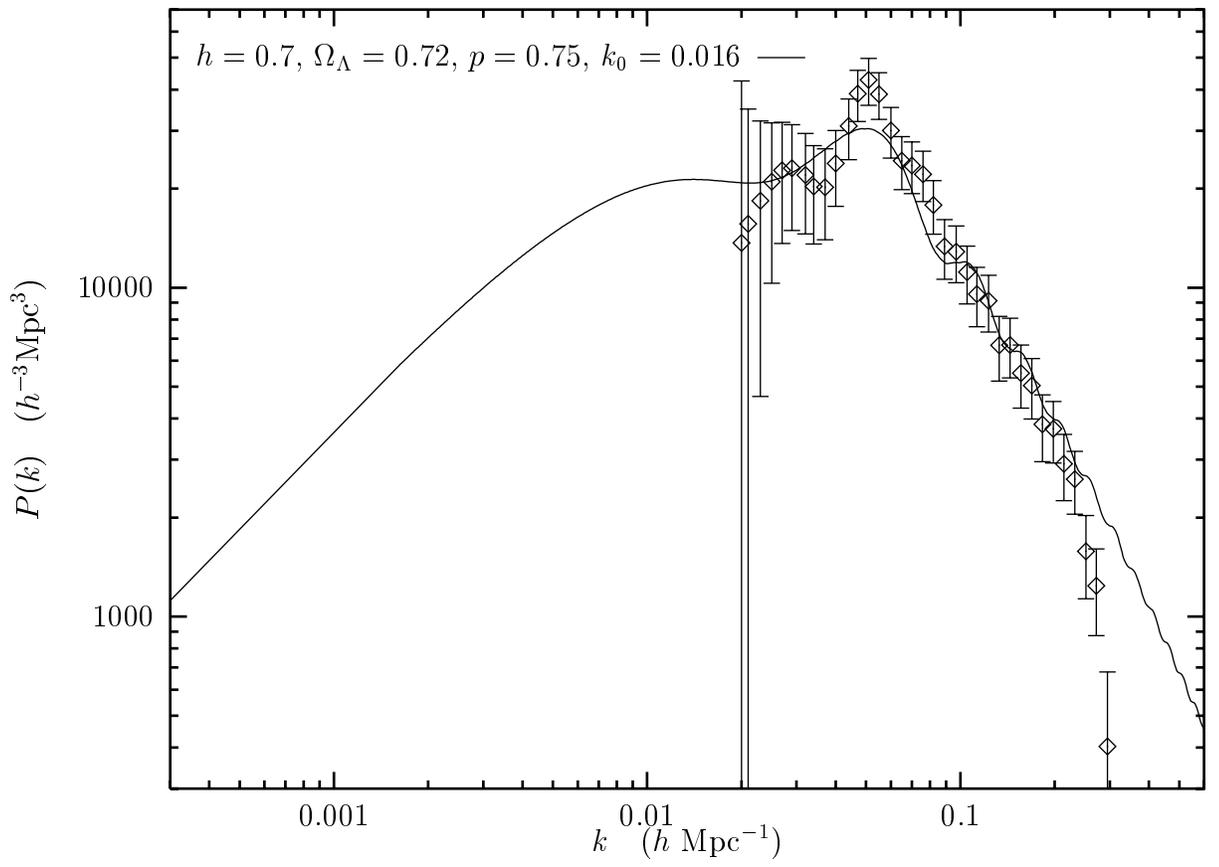
\end{document}